\documentclass[useAMS,usenatbib]{mn2e}
\usepackage{graphicx}
\usepackage{rotfloat}

\newcommand{\ms}{$\,$M$_\mathrm{\odot}$}

\newcommand{\be}{\begin{equation}}
\newcommand{\ee}{\end{equation}}
\newcommand{\stars}{{\sc stars}}
\newcommand{\el}[2]{\ensuremath{^{#1}\mathrm{#2}}}

\title[Modelling CEMP RR Lyraes]{Modelling the nucleosynthetic properties of carbon-enhanced metal-poor RR Lyrae stars}
\author[R.~J. Stancliffe et al.]{Richard J. Stancliffe$^{1,2}$\thanks{E-mail:
rjstancl@astro.uni-bonn.de}, Catherine R. Kennedy$^2$, Herbert H.~B. Lau$^1$ and 
\newauthor Timothy C. Beers$^{3,4}$\\
$^1$Argelander-Institut f\"ur Astronomie, Auf dem H\"ugel 71, 53121 Bonn, Germany \\
$^2$Research School of Astronomy \& Astrophysics, Australian National University, Canberra, ACT 2611, Australia \\
$^3$National Optical Astronomy Observatory, Tucson, AZ 85719, U.S.A.  \\
$^4$JINA, Joint Institute for Nuclear Astrophysics, East Lansing, MI 48824, U.S.A.
}
\begin{document}
\bibliographystyle{mn2e}

\date{Accepted 0000 December 00. Received 0000 December 00; in original form 0000 October 00}

\pagerange{\pageref{firstpage}--\pageref{lastpage}} \pubyear{0000}

\maketitle

\label{firstpage}

\begin{abstract}

Certain carbon-enhanced metal-poor stars likely obtained their composition via pollution from some of the earliest generations of asymptotic giant branch stars and as such provide important clues to early Universe nucleosynthesis. Recently, Kinman et al. discovered that the highly carbon- and barium-enriched metal-poor star SDSS~J1707+58 is in fact an RR Lyrae pulsator. This gives us an object in a definite evolutionary state where the effects of dilution of material during the Main Sequence are minimised owing to the object having passed through first dredge-up. We perform detailed stellar modelling of putative progenitor systems in which we accreted material from asymptotic giant branch stars in the mass range 1-2\ms. We investigate how the surface abundances are affected by the inclusion of mechanisms like thermohaline mixing and gravitational settling. While we are able to find a reasonable fit to the carbon and sodium abundances of SDSS~J1707+58, suggesting accretion of around 0.1\ms\ from a 2\ms\ companion, the strontium and barium abundances remain problematic and this object may have experienced something other than a standard $s$ process. We have more success in fitting the abundances of the mildly carbon-enriched, metal-poor RR Lyrae pulsator TY Gru (CS 22881-071), which we suggest received  0.1\ms\ of material from a companion of around 1\ms.
\end{abstract}


\section{Introduction}

Carbon-enhanced metal-poor (CEMP) stars with enrichments in $s$-process elements are believed to form via mass transfer in binary systems which start off with orbital periods of the order of thousands of days. The initially more massive star in the system (the primary) is assumed to be in the mass range 1-3\ms, while the initially less massive star (the secondary) would have a mass of less than 0.8\ms. The primary evolves to the thermally-pulsing asymptotic giant branch (TP-AGB) where it synthesises carbon and $s$-process elements \citep[see, e.g., the review by][for details]{2005ARA&A..43..435H} and brings these elements to the surface. As the star's envelope is stripped by the stellar wind, some of this ejecta material falls on to the surface of the secondary forming a carbon- and $s$-element-enriched object.

Predictions of the surface composition of these objects depends on what happens to the accreted material. In the classical stellar evolution picture, where only convective mixing is taken into account, the accreted material remains at the secondary's surface until the star leaves the main sequence, because low-metallicity, low-mass stars do not have substantial convective envelopes and there is no mechanism to transport material from the stellar surface. When the star begins to ascend the giant branch, a deep convective envelope begins to develop and the accreted material becomes mixed into the stellar interior. One would therefore expect red giant stars that have accreted material from a companion to have lower surface abundances for AGB-synthesised material than their less-evolved counterparts.

However, the above picture neglects an important piece of physics. The material ejected by the AGB star has undergone nucleosynthesis and is consequently of a higher mean molecular weight than the pristine surface material of the secondary. The situation of higher mean molecular weight material lying on top of lower mean molecular weight material is secularly unstable. The layers become mixed via the process of thermohaline convection\footnote{The terms `thermohaline mixing' and `thermohaline convection' are used interchangeably throughout.}, a process that has been recognized as important in mass-transfer binary systems \citep*[e.g.][]{2003NewA....8...23B}. In the context of CEMP stars, \citet{2007A&A...464L..57S} showed that thermohaline mixing could dramatically alter the surface compositions of the secondary after accretion. In their case of accretion of around  0.1\ms\ of material from a 2\ms\ companion, they found that within a tenth of the main sequence lifetime the accreted material was mixed throughout 90 percent of the star. Mixing of this efficiency means there is little dilution of accreted material when the convective envelope deepens as the star ascends the giant branch and there would be less difference between the surface abundances of such main-sequence and giant stars.

Thermohaline mixing is not the only non-convective process that may affect the surface abundances of CEMP stars. For example, gravitational settling can inhibit the action of thermohaline mixing \citep{2008ApJ...677..556T, 2009MNRAS.394.1051S}. Accreted material may also be mixed into the secondary by the action of mechanisms associated with rotation \citep{2012ApJ...751...14M}. The extent to which accreted material becomes mixed -- by whatever process(es) -- with the pristine material of the secondary is difficult to determine, but an average depth of mixed material of around 0.2\ms\ may be required \citep{2008ApJ...679.1541D}. For this reason, it is problematic to use the abundances of subgiant CEMP stars with $s$-process enhancements (CEMP-$s$ stars) as probes of AGB nucleosynthesis at low metallicity; one cannot be sure of how much dilution the accreted material has undergone \citep[for example, note the range of dilution factors used by][]{2012MNRAS.422..849B}.

Regardless of what happens to accreted material when the secondary is on the main sequence, when the star reaches the giant branch the development of a deep convective envelope (the cause of first dredge-up, FDU) will certainly affect the star's surface composition. A 0.8\ms\ star with a metallicity of [Fe/H] = -2.3 has an envelope that contains nearly 0.5\ms\ of material at its deepest extent \citep[e.g.][]{2008MNRAS.389.1828S}. This depth is not substantially affected by the composition or quantity of accreted material. In the absence of dilution by any other process, the action of convection on the giant branch will lead to the dilution of accreted material. This is often overlooked in many studies of CEMP-$s$ star abundances. Furthermore, because the depth of FDU is almost independent of envelope composition (for a given metallicity and a reasonable range of composition of accreted material) for stars of around 0.8\ms, we have a good handle on the extent of dilution any object that is more evolved than the end of FDU.

Often, CEMP stars in the giant stage are observed for detailed abundance analysis.  These cooler-temperature stars typically have significant amounts of molecular absorption in their spectra, making CNO abundance estimates achievable even at moderate resolution (e.g. C + N abundances from CH, CN, NH in optical spectra and O abundances from CO in near-IR spectra, \citealt{2011AJ....141..102K}).  However, for these cooler stars, one of the largest sources of error in abundance determinations arises from the difficulty of continuum placement amidst such molecular bands.  By observing stars with higher effective temperatures in the RR Lyrae phase, we minimize these uncertainties, though we may not be able to obtain as complete a chemical inventory as for cooler stars. From the point of view of stellar evolution theory, the RR Lyrae phase has the advantage that we observe the {\it integrated} effect of any mixing on the giant branch, including that taking place on the upper giant branch (the cause of which is uncertain, as outlined in more detail below).

The variable nature of RR Lyrae stars poses some challenges for carrying out spectroscopic observations. The radial pulsations of these stars mean that their surface gravity and effective temperature will significantly vary over short periods of time. These pulsations also impact the star's spectral lines, causing them to shift due to the radial velocities of the gas in the outer layers of the star; these radial velocities can vary by up to 70 km/s during a pulsation cycle. The effects of the radial pulsations can be corrected for if the phase in the star's pulsation cycle is known. The radial velocities of the star's outer layers are lowest when the star is at minimum light \citep{2012AJ....144...88H}. RR Lyrae stars should be observed when their phase is $\phi$=0.2-0.8 \citep[see Fig.~1 from ][]{2012ApJ...755L..18K} where maximum light occurs at $\phi$=0 and $\phi$=1. This range of phase corresponds to the minimum light, when an RR Lyrae star is least active.

The star SDSS J170733.93+585059.7 (hereafter SDSS~J1707+58) was originally thought to be a binary system on the basis of strong radial velocity variations \citep{2008ApJ...678.1351A}. However, recent work by \citet{2012ApJ...755L..18K} suggests that these variations are in fact due to stellar pulsations and that the star is an RR Lyrae pulsator. This places the object in a definite evolutionary stage, namely core helium-burning, and any material the star received from a binary companion would definitely have undergone (some) dilution by this post giant branch stage. \citet{2012ApJ...755L..18K} derive a metallicity of [Fe/H] = -2.92 for this object and their determination of the carbon abundance, namely  [C/Fe] = +2.79, makes SDSS~J1707+58 a highly carbon-rich star. In addition, they determine several other element abundances (see their table 1). Of particular interest are their measurements of strontium and barium, [Sr/Fe] = +0.75 and [Ba/Fe] = +2.83.

One other (relatively) carbon-rich RR Lyrae star is known, namely TY Gru (also identified as CS 22881-071 in the HK survey of \citealt{1992AJ....103.1987B}). \citet{2006AJ....132.1714P} report that this object has [C/Fe] = +0.89. Many authors define CEMP stars according to the definition of \citet{2005ARA&A..43..531B}, namely:  [Fe/H]$\ < -1.0$ and [C/Fe]$\ > +1.0$.  However, recent studies that include large samples of metal-poor stars have revealed that a more natural separation between carbon-normal and carbon-enhanced stars lies at the lower [C/Fe] value of +0.7 (see, for example, Figure 4 in \citealt{2012ApJ...744..195C}).  We therefore accept TY Gru as a CEMP star. TY Gru has a much more extensive set of abundance determinations than SDSS~J1707+58. In particular, \citet{2006AJ....132.1714P} give abundance measurements for the light $s$-process elements Sr, Y and Zr, the heavy $s$-elements, Ba, Ce and La, as well as for Pb. More recent abundance determinations for a limited subset of these elements have also been made by \citet{2011ApJS..197...29F}. Both groups agree that the object has substantial $s$-process enhancement, thus we should consider this object in light of the scenario of mass transfer from an AGB companion.

In this paper, we use detailed stellar evolution models together with up-to-date $s$-process nucleosynthesis calculations to study the surface abundances of low-mass stars during the RR Lyrae phase. We consider the accretion of material from a range of AGB companions, and study the effects of various assumptions about the physics affecting how accreted material is mixed into the secondary. These models are then compared to the observed abundances of the two known C-rich RR Lyrae stars.

\section{The stellar evolution code}

Calculations in this work have been carried out using a modified version of the \stars\ stellar evolution code originally developed by \citet{1971MNRAS.151..351E} and updated by many authors \citep[e.g.][]{1995MNRAS.274..964P, 2009MNRAS.396.1699S}. This code solves the equations of stellar structure and chemical evolution in a fully simultaneous manner, iterating on all variables at the same time in order to converge a model \citep[see][for a detailed discussion]{2006MNRAS.370.1817S}. The version used here includes the nucleosynthesis routines of \citet{2005MNRAS.360..375S} and \citet{stancliffe05}, which follow the nucleosynthesis of 40 isotopes from D to \el{32}{S} and important iron group elements. To this suite of isotopes, we have added the additional elements (without distinguishing individual isotopes, because observed isotopic ratios are not currently available): Sr, Y, Zr, Ba, La, Ce, Eu and Pb. These elements are only affected by the various mixing processes in the code and do not take part in any nuclear reactions. This is a reasonable approximation because these heavy nuclei are only affected by neutron captures and there is no active neutron source in the phases of evolution under consideration (however, see the Discussion regarding the possible occurrence of proton ingestion episodes).

The modelling of the accretion phase is performed in the manner described by \citet{2008MNRAS.389.1828S} and \citet{2009MNRAS.394.1051S}. Accretion begins at an age appropriate to the end of the life of the mass donor AGB star. Accretion is assumed to take place at a rate of $10^{-6}$\ms/yr, a value consistent with accretion via the \citet{1944MNRAS.104..273B} prescription assuming the AGB star is in its superwind phase\footnote{This may be an underestimate as wind mass transfer is considerably more complicated. The interested reader is directed to \citet{2013A&A...552A..26A} for a detailed discussion.} and this accretion takes place at a time consistent with the onset of this superwind in the AGB star. The yields for the AGB stars are taken from \citet{2012ApJ...747....2L} (see \citealt{2010MNRAS.403.1413K} for a discussion of the stellar evolution code used) and a summary of the composition of the accreted ejecta is given in Table~\ref{tab:yields}. As we are unable to distinguish between the various isotopes of the heavy elements, we have added together the total mass of all the isotopes of a given element in order to calculate the yield. In each case accretion is assumed to continue until the final mass of the accreting star is 0.8\ms. We have started with initial stellar masses of 0.7, 0.79 and 0.799\ms, corresponding to accreting 0.1, 0.01 and 0.001\ms\ of material respectively, in each case giving an object whose post-accretion mass is 0.8\ms. For each of these cases, we accrete material from three different masses of donor AGB star, namely: 1, 1.5 and 2\ms. Once accretion is completed we employ a \citet{1975MSRSL...8..369R} mass-loss prescription with $\eta=0.4$ throughout the remaining evolution. Evolution is terminated at the beginning of the thermally-pulsing AGB phase. All models have a metallicity of $Z=10^{-4}$, corresponding to [Fe/H]$ = -2.3$.

\begin{table}
\begin{center}
\begin{tabular}{cccccc}
Donor Mass & [C/Fe] & [Sr/Fe] & [Ba/Fe] & [Eu/Fe] & [Pb/Fe] \\
(\ms) & \\ 
\hline
1 & 1.74 & 0.97 & 1.89 & 0.99 & 2.32 \\
1.5 & 2.88 & 1.62 & 2.43 & 1.45 & 3.14 \\
2 & 3.19 & 1.77 & 2.40 & 1.50 & 3.23 \\
\hline
\end{tabular}
\end{center}
\caption{Composition of material accreted from the donor star.}
\label{tab:yields}
\end{table}

We have run models with three assumptions about the physics of the accretion process. The simplest case is where material is only allowed to mix via convection. In the next case, we assume that thermohaline mixing is active. Thermohaline mixing is implemented via the diffusive prescription outlined in \citet{1980A&A....91..175K}. In the final case, we assume that both gravitational settling and thermohaline mixing occur. In each case, the various physical mechanisms are assumed to be active throughout the run so that they remain physically consistent. 

We have accounted for abundance changes on the upper part of the giant branch using a diffusive prescription for thermohaline mixing as described in \citet{1972ApJ...172..165U} and \citet{1980A&A....91..175K}. This mixing is only switched on at the end of FDU. Following \citet{2007A&A...467L..15C}, we set the free parameter of the diffusion coefficient to 1000, a value which these authors have shown reproduces the abundance changes in field stars. The same value has also been shown to reproduce the abundance changes in both carbon-rich and carbon-normal metal-poor stars \citep{2009MNRAS.396.2313S}, as well as for globular cluster stars \citep{2012ApJ...749..128A,2011ApJ...728...79A}. However, it should be noted that hydrodynamical simulations do not support such a large value for the diffusion coefficient \citep{2011ApJ...728L..29T,2011ApJ...727L...8D}. Hence we cannot be certain that thermohaline mixing is indeed responsible for abundance changes on the upper giant branch. However, as abundance changes are only observed to occur for the light elements up to nitrogen \citep[see e.g.][]{2000A&A...354..169G} this uncertainty does not affect our predictions for the heavy elements, particularly the $s$-process elements.

\section{Results}

\begin{figure}
\includegraphics[width=\columnwidth]{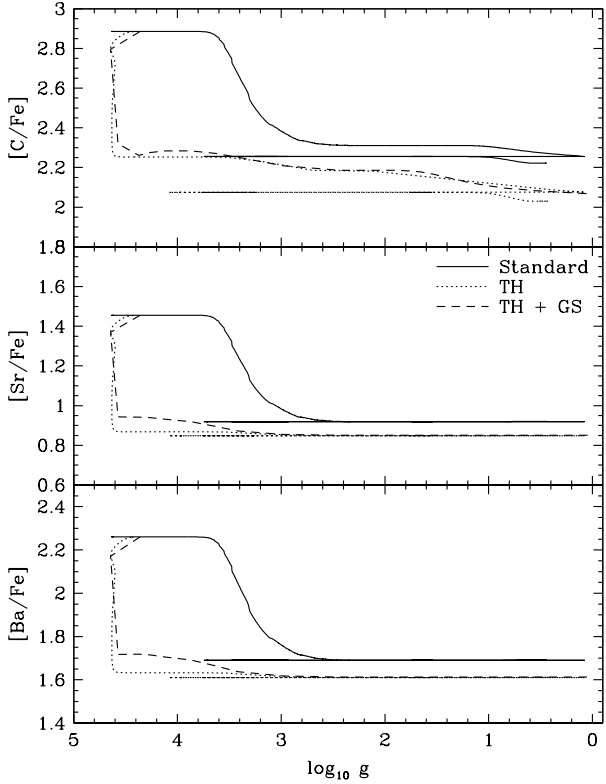}
\caption{[C/Fe], [Sr/Fe] and [Ba/Fe] as a function of $\log g$ under the assumption of various combinations of mixing mechanism: convection only (solid line), thermohaline mixing and convection (dotted line) and thermohaline mixing, gravitational settling and convection (dashed line). In each case, 0.1\ms\ of material has been accreted from a 1.5\ms\ companion.}
\label{fig:mixing}
\end{figure}

\begin{figure}
\includegraphics[width=\columnwidth]{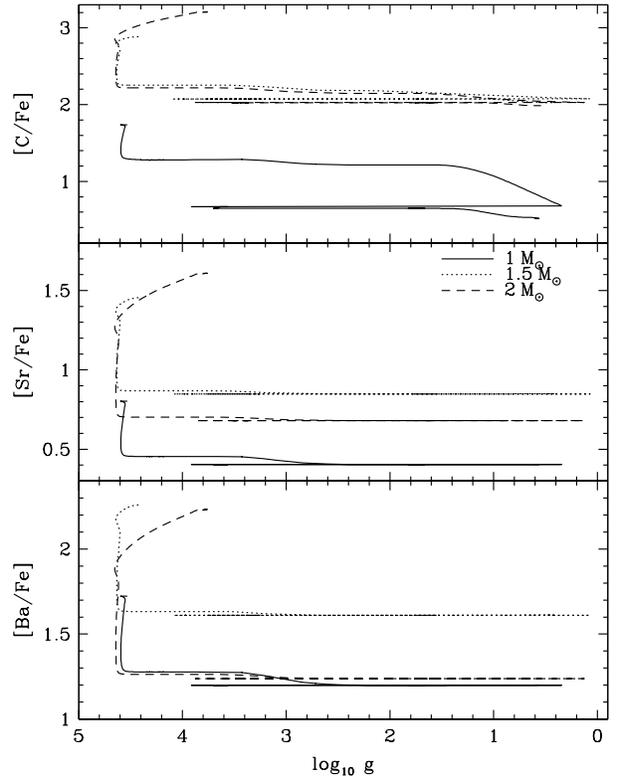}
\caption{[C/Fe], [Sr/Fe] and [Ba/Fe] as a function of $\log g$ for the case of accretion of 0.1\ms\ of material from companions of 1- (solid line), 1.5- (dotted line) and 2-\ms\ (dashed line) companion. For each model, only convection and thermohaline mixing are included.}
\label{fig:masses}
\end{figure}

\begin{figure}
\includegraphics[width=\columnwidth]{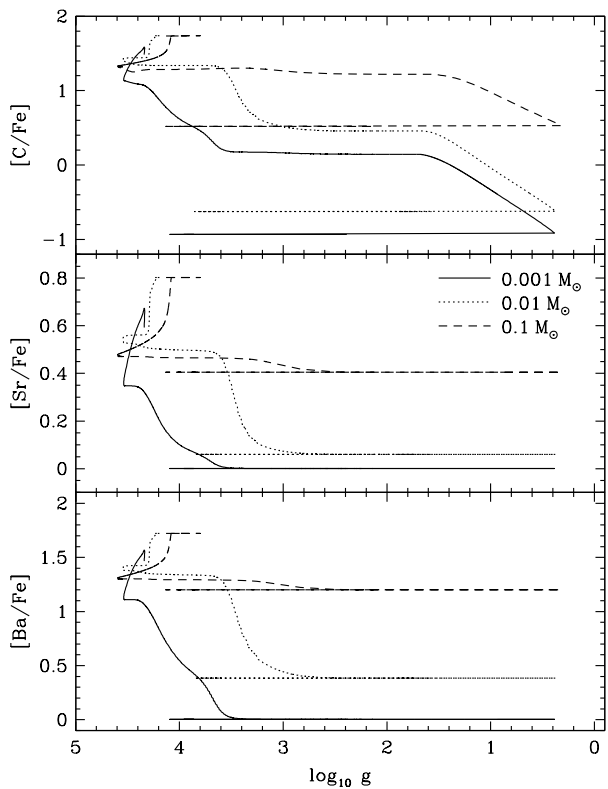}
\caption{[C/Fe], [Sr/Fe] and [Ba/Fe] as a function of $\log g$ for the case of accretion of 0.001 (solid line), 0.01 (dotted line) and 0.1\ms\ (dashed line) of material from a 1\ms\ companion. For each model, convection, thermohaline mixing and graviational settling are all included.}
\label{fig:accretion}
\end{figure}

The surface abundances for select elements of our models during the core He-burning phase are given in Tables~\ref{tab:abunds_1sm}, \ref{tab:abunds_1.5sm} and \ref{tab:abunds_2sm}. These abundances remain constant throughout the core helium burning phase. As may be expected, we find that the amount of material accreted is the singularly most important quantity in determining the abundances during core He burning. In each case, regardless of the mixing physics employed, the models in which 0.1\ms\ of material was accreted have the most enriched surfaces as expected. Figures~\ref{fig:mixing}, \ref{fig:masses} and \ref{fig:accretion} show the variation of [C/Fe], [Sr/Fe] and [Ba/Fe] as a function of $\log g$ for a selection of the models.

\begin{table*}
\begin{center}
\begin{tabular}{cccccccccc}
Accreted Mass & [C/Fe] & [Sr/Fe] & [Y/Fe] & [Zr/Fe] & [Ba/Fe] & [La/Fe] & [Ce/Fe] & [Eu/Fe] & [Pb/Fe] \\
(\ms) & \\ 
\hline
\multicolumn{10}{c}{Standard -- convection alone} \\
\hline
0.1 & 0.72 & 0.39 & 0.49 & 0.68 & 1.17 & 1.17 & 1.30 & 0.40 & 1.60 \\
0.01 & -0.28 & 0.05 & 0.07 & 0.11 & 0.32 & 0.32 & 0.39 & 0.05 & 0.60 \\
0.001 & -0.55 & 0.00 & 0.00 & 0.00 & 0.00 & 0.00 & 0.00 & 0.00 & 0.00 \\
\hline
\multicolumn{10}{c}{Thermohaline mixing, no gravitational settling} \\
\hline
0.1 & 0.65 & 0.40 & 0.51 & 0.70 & 1.20 & 1.20 & 1.31 & 0.41 & 1.62 \\
0.01 & -0.23 & 0.07 & 0.10 & 0.17 & 0.43 & 0.43 & 0.52 & 0.08 & 0.76 \\
0.001 & -0.52 & 0.00 & 0.00 & 0.00 & 0.02 & 0.02 & 0.03 & 0.00 & 0.05 \\
\hline
\multicolumn{10}{c}{Thermohaline mixing and gravitational settling} \\
\hline
0.1 & 0.52 & 0.40 & 0.51 & 0.70 & 1.20 & 1.20 & 1.31 & 0.41 & 1.62 \\
0.01 & -0.62 & 0.06 & 0.08 & 0.14 & 0.38 & 0.38 & 0.46 & 0.06 & 0.67 \\
0.001 & -0.93 & 0.00 & 0.00 & 0.00 & 0.00 & 0.00 & 0.00 & 0.00 & 0.01 \\
\hline
\end{tabular}
\end{center}
\caption{Surface abundances of models for stars in the horizontal-branch stage of evolution assuming  a 1\ms\ donor.}
\label{tab:abunds_1sm}
\end{table*}

For accretion from the 1\ms\ companion, we find very similar core He-burning surface abundances regardless of the treatment of the mixing physics, as can be seen in Fig.~\ref{fig:mixing}. This is because the mixing of accreted material is dominated by the occurrence of FDU. The 1\ms\ model does not undergo much enrichment through third dredge-up (note that [C/Fe] is over 1 dex lower than the other models, as can be seen in Fig.~\ref{fig:masses}). Consequently, the accreted material does not have a particularly high mean molecular weight which means that it does not mix deeply (via thermohaline convection) while the star is on the main sequence. Non-convective mixing reaches to depths of around 0.4\ms\ in the most extreme case, which is crucially {\it shallower} than the depth reached by the convective envelope during FDU. In essence, after FDU the three models have exactly the same dilution factor and this is fixed by the maximum depth of the convective envelope. In such a case, we can confidently predict the composition of the AGB ejecta given the observed surface abundances of a CEMP RR-Lyrae star. The only exception is for the carbon abundance, which shows some variation. These changes are related to the processing of carbon by mixing on the upper part of the red giant branch.

The carbon abundance deserves some comment. Gravitational settling, in certain limited circumstances,  produces a molecular weight barrier that inhibits the action of thermohaline mixing. Na\"ively, one might expect that the models with thermohaline mixing alone might exhibit the lowest carbon abundances. In this case, the dilution of accreted material is maximised. In the cases of low accretion mass, where the depth of mixing is less than the depth to which convection reaches during FDU, one would expect the abundances to be equal as dilution is dominated by this event which is equal in both cases. However, we find that this is not the case. The reason is as follows. In these models we have included mass loss \citep[which was not included by][]{2008MNRAS.389.1828S}. By the time one of these models reaches the base of the giant branch (and the onset of FDU) around 0.004\ms\ of material has been removed from the surface. This is an insignificant amount to affect the subsequent evolution, but it {\it does} affect the chemical abundances. In the cases with gravitational settling, more of the accreted material remains at the surface from whence it is ablated by the stellar wind. By the onset of FDU there is less accreted material left to be diluted in the convective envelope. This explains why the thermohaline mixing only models display greater abundances that those including both thermohaline mixing and gravitational settling. In the case of large amounts of accretion, gravitational settling is ineffective in preventing thermohaline mixing because the quantity of matter easily overwhelms the molecular weight barrier \citep[see][for further details]{2008MNRAS.389.1828S} while in the case that only a small quantity of material is accreted, dilution at FDU dominates the mixing.

\begin{table*}
\begin{center}
\begin{tabular}{cccccccccc}
Accreted Mass & [C/Fe] & [Sr/Fe] & [Y/Fe] & [Zr/Fe] & [Ba/Fe] & [La/Fe] & [Ce/Fe] & [Eu/Fe] & [Pb/Fe] \\
(\ms) & \\ 
\hline
\multicolumn{10}{c}{Standard -- convection alone} \\
\hline
0.1 & 2.26 & 0.92 & 1.03 & 1.19 & 1.68 & 1.70 & 1.77 & 0.77 & 2.39 \\
0.01 & 0.52 & 0.15 & 0.18 & 0.26 & 0.57 & 0.56 & 0.63 & 0.10 & 1.16 \\
0.001 & -0.58 & 0.00 & 0.00 & 0.00 & 0.00 & 0.00 & 0.00 & 0.00 & 0.03 \\
\hline
\multicolumn{10}{c}{Thermohaline mixing, no gravitational settling} \\
\hline
0.1 & 2.07 & 0.85 & 0.95 & 1.12 & 1.61 & 1.62 & 1.70 & 0.70 & 2.31 \\
0.01 & 0.48 & 0.15 & 0.19 & 0.26 & 0.56 & 0.57 & 0.63 & 0.10 & 1.16 \\
0.001 & -0.44 & 0.01 & 0.02 & 0.02 & 0.06 & 0.06 & 0.08 & 0.01 & 0.26 \\
\hline
\multicolumn{10}{c}{Thermohaline mixing and gravitational settling} \\
\hline
0.1 & 2.07 & 0.85 & 0.95 & 1.12 & 1.61 & 1.62 & 1.70 & 0.70 & 2.31 \\
0.01 & 0.39 & 0.13 & 0.17 & 0.23 & 0.52 & 0.53 & 0.59 & 0.09 & 1.11 \\
0.001 & -0.57 & 0.00 & 0.00 & 0.00 & 0.00 & 0.01 & 0.01 & 0.00 & 0.06 \\
\hline
\end{tabular}
\end{center}
\caption{Surface abundances of models for stars in the horizontal-branch stage of evolution, assuming a 1.5\ms\ donor.}
\label{tab:abunds_1.5sm}
\end{table*}

For the case of accretion from a 1.5 or 2\ms\ companion we are able to draw some distinctions between our various mixing scenarios. For both companion masses, the standard convective case produces the greatest surface enrichments at the RR Lyrae phase. If 0.1\ms\ is accreted from a 1.5\ms\ companion, the [C/Fe] abundance is 0.2 dex higher for the standard case than for the other two mixing cases. If the companion is 2\ms, this figure increases to 0.5 dex. The heavier elements are less affected (they do not undergo additional nuclear processing, unlike carbon), and changes of around 0.08 dex and 0.4 dex are found for the 1.5 and 2\ms\ companions. In both cases, we are unable to distinguish between the thermohaline mixing model and the thermohaline mixing plus gravitational settling case. This is also true for accretion of 0.01 or 0.001\ms\ of material from either companion.

\begin{table*}
\begin{center}
\begin{tabular}{cccccccccc}
Accreted Mass & [C/Fe] & [Sr/Fe] & [Y/Fe] & [Zr/Fe] & [Ba/Fe] & [La/Fe] & [Ce/Fe] & [Eu/Fe] & [Pb/Fe] \\
(\ms) & \\ 
\hline
\multicolumn{10}{c}{Standard -- convection alone} \\
\hline
0.1 & 2.58 & 1.04 & 1.11 & 1.22 & 1.64 & 1.67 & 1.80 & 0.79 & 2.47 \\
0.01 & 0.95 & 0.20 & 0.24 & 0.29 & 0.55 & 0.58 & 0.67 & 0.12 & 1.28 \\
0.001 & -0.52 & 0.00 & 0.00 & 0.00 & 0.00 & 0.00 & 0.00 & 0.00 & 0.07 \\
\hline
\multicolumn{10}{c}{Thermohaline mixing, no gravitational settling} \\
\hline
0.1 & 2.02 & 0.68 & 0.74 & 0.84 & 1.24 & 1.27 & 1.39 & 0.47 & 2.06 \\
0.01 & 0.62 & 0.14 & 0.16 & 0.20 & 0.41 & 0.43 & 0.51 & 0.08 & 1.07 \\
0.001 & -0.36 & 0.01 & 0.02 & 0.02 & 0.05 & 0.05 & 0.06 & 0.01 & 0.24 \\
\hline
\multicolumn{10}{c}{Thermohaline mixing and gravitational settling} \\
\hline
0.1 & 2.04 & 0.69 & 0.75 & 0.85 & 1.25 & 1.28 & 1.39 & 0.47 & 2.07 \\
0.01 & 0.61 & 0.13 & 0.16 & 0.19 & 0.40 & 0.42 & 0.51 & 0.07 & 1.06 \\
0.001 & -0.42 & 0.00 & 0.00 & 0.00 & 0.00 & 0.00 & 0.00 & 0.00 & 0.07 \\
\hline
\end{tabular}
\end{center}
\caption{Surface abundances of models for stars in the horizontal-branch stage of evolution, assuming  a 2\ms\ donor.}
\label{tab:abunds_2sm}
\end{table*}

\begin{figure}
\includegraphics[width=\columnwidth]{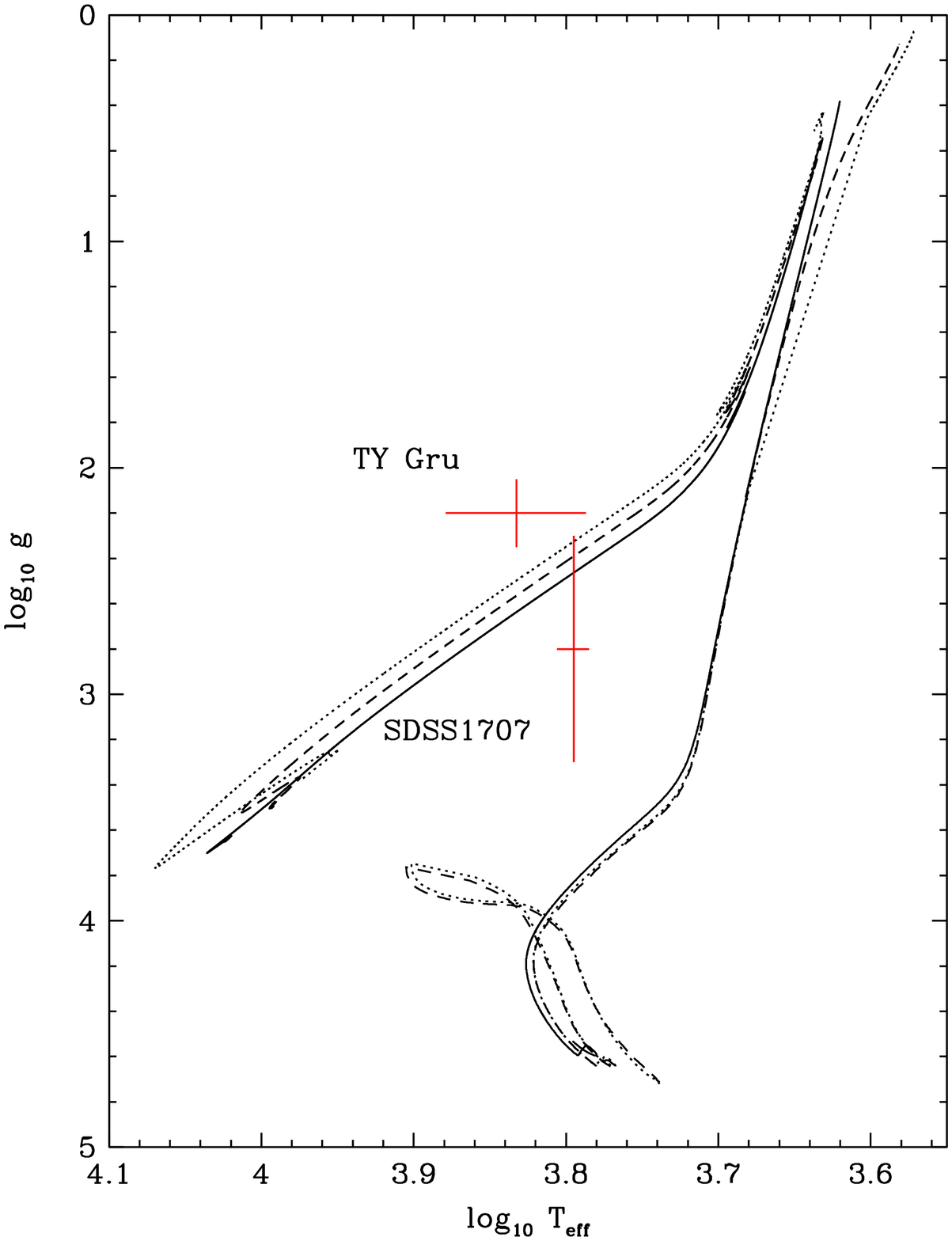}
\caption{Evolutionary tracks in $\log_{10} T_\mathrm{eff}$-$\log_{10} g$ space. The models shown are for the case when 0.1\ms\ of material is accreted from a 1- (solid line), 1.5- (dotted line) and 2\ms\ (dashed line) companion, with only convection and thermohaline mixing accounted for. These tracks show the evolution from the zero-age main sequence to the early AGB phase. There is a break in the tracks between helium ignition at the tip of the red giant branch and the onset of quiescent helium burning. The locations of SDSS~J1707+58 and TY Gru are also displayed.}
\label{fig:TG_masses}
\end{figure}

\section{Discussion}

We now discuss these models in the context of the two carbon-enhanced RR Lyrae stars currently known: SDSS~J1707+58 and TY Gru.

\subsection{SDSS~J1707+58}

In Figure~\ref{fig:TG_masses}, we plot the evolutionary tracks for three of our models in the $\log T_\mathrm{eff} - \log g$ plane. These tracks follow the evolution from the end of accretion to the beginning of the thermally pulsing AGB phase. While our tracks do pass through the errors on the $\log g$ determination, they are toward the low end. This suggests that SDSS~J1707+58 may be slightly more massive and/or compact than the models presented here. Variations in accretion mass, companion mass and mixing physics shift the tracks by no more than 0.2 dex in  $\log g$.

SDSS~J1707+58 presents some problems for fitting to the models we have produced. First, its [Fe/H] is about 0.6 dex lower than our input models. Assuming there is no change in the behaviour of the donor AGB star (i.e., there is no  occurrence of proton ingestion episodes (PIEs) or other events that would affect the nucleosynthesis -- we return to the subject of these events below), we may feel justified in increasing the computed carbon abundance by 0.6 dex to account for the missing iron. This would bring the carbon abundance within reach of the stellar models, but we are forced to assume the AGB donor is at least 1.5\ms. We also conclude that the amount of material that was accreted must also have been around 0.1\ms, as anything less would produce too low a value for [C/Fe] at this stage in the evolution.

Unfortunately, it is difficult to reconcile the observed heavy element abundances with the models presented here. The measured barium abundance is considerably higher than anything we have in our models. At best, we have [Ba/Fe] = +1.68. In fact, the observed abundance of [Ba/Fe] = +2.83 is higher than anything we produce in the AGB donor models before any dilution into the secondary is taken into account. At the same time, our most barium-enhanced model shows more strontium than is observed. The observed abundance is [Sr/Fe] = +0.75, whereas our models tend to exhibit [Sr/Fe] in excess of +1 prior to mixing. The unusually high [Ba/Sr] = +2.08 may suggest that this object's heavy elements do not come from a standard $s$-process.  An [hs/ls] of around +2 is more common for objects in the CEMP-r/s class, rather than the CEMP-s class\footnote{CEMP stars are usually divided into subgroups based on their heavy element content. Here we have followed the definitions given by \citet{2010A&A...509A..93M}, in which CEMP-s stars have [Ba/Fe]$>+1$ and CEMP-r/s stars have [Eu/Fe]$>+1$ and [Ba/Eu]$>0$ (i.e. they are still barium-enriched). Note that this defintion is slightly different from that given by \citet{2005ARA&A..43..531B}.}. The high [Ba/Fe] value is also more common for CEMP-r/s objects -- in fact, it would even be one of the more enriched objects in this class \citep[see e.g. figure 2 in ][]{2010A&A...509A..93M}. Should europium ever be measured in this object, we would expect it to be highly elevated, probably at the level of [Eu/Fe]$\approx+2$ based on the observed trends of CEMP-r/s stars.

Regarding the high barium abundance, we can make a simple estimate of the required abundance in the accreted ejecta from a putative AGB star companion if we make some assessment of the extent of mixing that has taken place. Heavy elements are not expected to undergo nuclear reactions during dilution so the total mass of a given element, $j$, within the star remains constant from the point that accretion has finished. In this case, we have:
\be
X_\mathrm{acc}\Delta M_\mathrm{acc} + X_\mathrm{i}\Delta M_\mathrm{mix} = X_\mathrm{f}\Delta M_\mathrm{f},
\ee
where $X$ is the abundance of a particular species, the subscripts `acc', `i' and `f' refer to the accreted, initial and final material respectively, and $\Delta M_\mathrm{acc}$ is the mass accreted, $\Delta M_\mathrm{mix}$ is the mass of pristine stellar material that will become mixed and $\Delta M_\mathrm{f}$ is the total mass over which mixing occurs (i.e. $\Delta M_\mathrm{mix} = \Delta M_\mathrm{f} - \Delta M_\mathrm{acc}$).
We need only specify a depth to which material is mixed, without reference to how this happens. It could be that material is mixed only by the deepening of the convective envelope, material could be efficiently mixed by thermohaline mixing (or any other process), or material may be partially mixed by thermohaline mixing and further mixed at FDU. The above formula only requires that we know the depth. From the stellar models, the maximum extent (in mass) of the convective envelope during the ascent of the giant branch is 0.5\ms.

In the case that the accreted material is significantly enriched in a given element compared to the pristine material of the secondary on to which it is placed, the above equation can be reduced to:
\be
X^j_\mathrm{acc} \approx X^j_\mathrm{f} {\Delta M_\mathrm{f}\over \Delta M_\mathrm{acc}}
\ee
where the log of the fraction is what some authors refer to as the dilution factor \citep[e.g.][]{2012MNRAS.422..849B}.

If we assume that the maximum depth of the convective envelope is the deepest that any accreted material gets mixed, we may derive the barium abundance required in the accreted material. If we assume that 0.1\ms\ of material was accreted, we obtain $X_\mathrm{Ba} = 8.54\times10^{-8}$ which is equivalent to [Ba/Fe] = +3.61. If the accreted material was only 0.001\ms, this value jumps by two orders of magnitude to [Ba/Fe] = +5.61. The former value is problematic for current stellar models; the latter is simply impossible. 

Finally, we have the issue of the sodium abundance. [Na/Fe] is highly elevated, being measured as +2.40. This value is reached by our model which accretes 0.1\ms\ of material from a 2\ms\ companion and where no thermohaline convection is considered. If we include thermohaline convection, the model reaches [Na/Fe] = +2.07, which is somewhat lower though perhaps not incompatible with the observed value. The 1.5\ms\ model produces much less sodium and only reaches [Na/Fe] = +1.78. While the Na abundance is large with respect to most of the models, it should be noted that the sodium lines used for the abundance analysis are known to be affected by NLTE effects.  The NLTE corrections can be quite large, reaching almost -1 dex in the most extreme cases \citep{2003ChJAA...3..316T}.  Assuming a lower Na abundance due to NLTE effects, we find that more combinations of progenitor mass and mixing processes could be consistent with the abundance. Note that \citet{2012MNRAS.422..849B} suggest that sodium can potentially be a discriminant of mass of the companion AGB star. For further details of sodium in the context of the AGB mass transfer scenario see \citet{2009MNRAS.394.1051S}.

In summary, it seems probable that SDSS~J1707+58 had a companion of around 2\ms\ from which it accreted at least 0.1\ms\ of material (possibly more\footnote{The maximum mass that such a system can accrete depends on how the mass transfer occurs and this is not well understood. We have assumed a \citet{1944MNRAS.104..273B} accretion law, for which the maximum accretion mass is likely to be around 0.1\ms. However, recent work by \citet{2013A&A...552A..26A} suggests that accretion in these systems could be more efficient and perhaps accretion masses of around 0.4\ms\ could be possible.}). However, the enrichments of strontium and barium are not well reproduced, perhaps because this object received material from a source that had not undergone a pure $s$-process. However, we cannot discount the possibility that large corrections due to NLTE effects, particularly for Sr II, could have a substantial effect on the interpretation of the progenitor of SDSS~J1707+58.  In a recent study of abundances in two EMP ([Fe/H]$< -3$) RR Lyrae stars, \citet{2011A&A...527A..65H} apply a correction to the Sr II abundance of +0.6 dex to RR Lyrae stars of similar atmospheric parameters following the prescription of \citet{2001A&A...376..232M} and \citet{1997ARep...41..530B}.  Applying a correction of this magnitude to the [Sr/Fe] abundance of SDSS~J1707+58, would bring the strontium abundance closer to the values obtained in the undiluted AGB ejecta. Further study of this object to determine other heavy element abundances, particularly europium and lead, would be highly desirable.

Because SDSS~J1707+58 has such a low metallicity, it is possible that the nucleosynthesis does not proceed in the same way as described in the models presented here. At low metallicity, helium-driven convective regions may be able to penetrate into hydrogen-rich regions in an event referred to as a proton ingestion episode (PIE). These can happen either during the core helium flash, or during the early helium shell flashes (thermal pulses) on the AGB \citep[e.g.][]{1990ApJ...349..580F}. A PIE in the AGB donor star would likely alter the composition of the accreted material, while a PIE in secondary at the core helium flash could also affect the surface composition. In both cases, heavy element abundances can be affected because proton ingestion leads to the formation of \el{13}{C} which can subsequently undergo alpha capture to release the neutrons required for heavy element nucleosynthesis \citep{2010A&A...522L...6C,2009PASA...26..139C}. However, the nucleosynthetic signature of these events is poorly understood because the 1D stellar evolution calculations do not accurately treat this phase. Hydrodynamical models of PIEs are now being developed and these will hopefully shed light on what happens in these events \citep{2011ApJ...742..121S,2011ApJ...727...89H}.

Current stellar models do not predict PIEs for stars at the metallicity of SDSS~J1707+58, either during the core helium flash or while the star is on the AGB  \citep[compare the calculations of][]{2008A&A...490..769C,2009MNRAS.396.1046L,2010MNRAS.405..177S}. However, there are many modelling uncertainties involved and we cannot rule out this possibility. For example, hydrodynamical simulations by \citet{2010A&A...520A.114M} suggest that these events could even occur in metal-rich stars.

\subsection{TY Gru}

\begin{table}
\begin{center}
\begin{tabular}{ccc}
Abundance &  \citet{2006AJ....132.1714P} & \citet{2011ApJS..197...29F} \\
\hline
 [CH/Fe] & 0.89$\pm$0.15 &  - \\
\,[Mg/Fe] & 0.37$\pm$0.10 & 0.38 \\
 \,[Fe/H] & -2.09$\pm$0.09 & -1.99 \\
 \,[Sr/Fe] & 0.60$\pm$0.22 & 0.04 \\
 \,[Y/Fe] & 0.26$\pm$0.05 & 0.43 \\
 \,[Zr/Fe] & 0.58$\pm$0.12 & 0.32 \\
 \,[Ba/Fe] & 1.23$\pm$0.15 & 1.05 \\
 \,[La/Fe] & 1.05$\pm$0.05 & 0.85 \\
 \,[Ce/Fe] & 1.05$\pm$0.15 &  - \\
 \,[Eu/Fe] &  0.69$\pm$0.05 & - \\
 \,[Pb/Fe] & 2.10$\pm$0.14 & - \\
 \hline
\end{tabular}
\end{center}
\caption{A selected list of surface abundances for TY Gru as determined by two separate groups.}
\label{tab:TYGru}
\end{table}

TY Gru has a metallicity that is about 0.3 dex higher than that of the models presented here, with \citet{2011ApJS..197...29F} giving [Fe/H] = -1.99 and \citet{2006AJ....132.1714P} giving a slightly lower value of [Fe/H] = -2.09. A list of some of the abundances determined for this star are provided in Table~\ref{tab:TYGru}. With an observed [C/Fe] of +0.89, few of the models presented here have a comparable carbon abundance. The 0.01\ms\ of accretion from a 2\ms\ companion comes closest. However, it is probable that if we were to accrete up to a few times as much material from either the 1.5 or 2\ms\ companions (with or without thermohaline mixing) we might also obtain an acceptable fit for the carbon abundance.

Turning our attention to the neutron-capture elements, we note that the object has [Ba/Fe] = +1.23 and [Eu/Fe] = +0.69, which places it among the CEMP-s stars. To reach this high an enrichment of barium, we are forced to invoke one of the 0.1\ms\ of accretion models, though these models tend to have a slightly higher barium abundance than observed. The europium abundance also gives a reasonable fit, though it is slightly under-abundant.

\begin{figure}
\includegraphics[angle=270,width=\columnwidth]{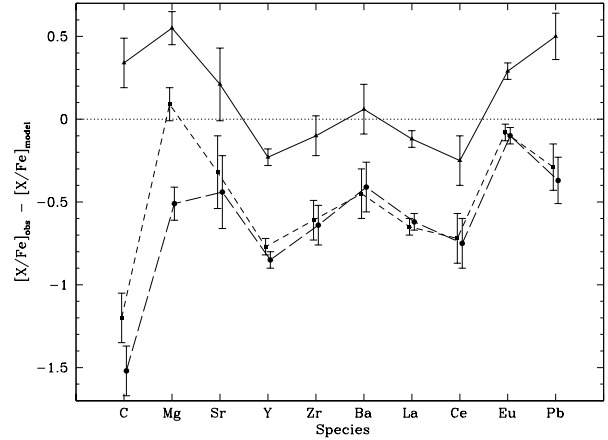}
\caption{Comparison of the observed abundances of TY Gru to those predicted by our models when accreting 0.1\ms\ from companions of 1 (triangles, solid lines), 1.5 (squares, short-dashed line) and 2\ms\ (circles, long-dashed line), together with errors on the observations. Positive values indicated the model is under-abundant; negative values indicate overabundance.}
\label{fig:residual}
\end{figure}

Rather than comparing individual abundances, we can try to fit all the abundances listed in Table~\ref{tab:TYGru} simultaneously. In Figure~\ref{fig:residual}, we plot the residual (observed abundance minus the model abundance) to such fits for three of our models to the \citet{2006AJ....132.1714P} data. These are all for accreting 0.1\ms\ of material from each of our companion masses and only including convective mixing (the analysis has been carried out for all our cases however). The best fit we are able to obtain is in the case that 0.1\ms\ of material is accreted from the 1\ms\ companion; as can be seen from Figure~\ref{fig:residual} it is barely a tolerable fit. In particular the yttrium and europium abundance in the model lie several sigma away from the observed values, with the former being overabundant and the latter under-abundant. The poor fit to the observed magnesium abundance can be easily reconciled as the models were solar-scaled and not alpha-enhanced. Typical observed alpha-enhancements are around 0.4 dex, and the measured [Mg/Fe] is consistent with this.

There is potentially an issue with the light s-element abundances. To within the error bars, the heavy s-elements have the same abundances. This is consistent with models of \citet{2012ApJ...747....2L} which show [Ba/Fe], [Ce/Fe] and [La/Fe] have the same abundances within 0.1 dex for all the masses considered (see their figure 2). However, the light s-elements show considerably more variation, with yttrium being less abundant that both strontium and zirconium. While there is considerably more variation in the relative abundances of Sr, Y and Zr predicted by the models of \citet{2012ApJ...747....2L}, none of their combinations show Y to be of lower abundance than both of its neighbouring elements. The same is true of the models of \citet{2010MNRAS.404.1529B} which are all computed at 1.5\ms\ but with a large variation in pocket size. Does this point to some deficiency in our understanding of the nuclear physics of the creation of the light s-elements, or is the lower Y abundance simply an observational artefact? This issue was also noted by \citet{2012MNRAS.422..849B}.

We still have a problem matching the model abundance to the light, heavy and Pb peaks. The models do not show the same relative peak heights as observed. If we fit the model so that the heavy-$s$ peak is reproduced, then we typically overproduce the light-$s$ peak while underproducing Pb. This could potentially point to problems with our neutron-capture nucleosynthesis. We reiterate that we have only looked at ejecta coming from stars in which we have included a partial mixing zone of $2\times10^{-3}$\ms. Our ignorance of the way the \el{13}{C} pocket is formed (including its shape and extent) may be hampering our comparison. \citet{2012ApJ...747....2L} show that if the partial mixing zone is varied in size, the ratios [Ba/Sr] and [Pb/Ba] can change quite dramatically, with the former varying between 0.55 and 0.81 and the latter between -0.32 and +0.69. Similarly, \citet{2010MNRAS.404.1529B} report that [Ba/Sr] and [Pb/Ba] vary between +0.46 and +1.13, and +0.88 and +2.44, respectively when they vary their standard pocket by a factor of 12. 

\section{Conclusion}

We have presented models for the surface composition of carbon-enriched, metal-poor stars during the core helium burning phase appropriate for those objects displaying RR Lyrae-like pulsations. This is done under the assumption that  such CEMP stars are formed by mass transfer in a binary system containing an AGB star. We have investigated a range of donor and accretion masses, and examined the effects of various mechanisms for the mixing of accreted material. We find that if enrichment is substantial enough, we are able to distinguish between the occurrence of these mixing mechanisms.

We have compared these predictions to the observed compositions of two known carbon-enhanced RR Lyrae pulsators: SDSS~J1707+58 and TY Gru. While we can reproduce the carbon and sodium abundances of SDSS~J1707+58, the object's barium abundance is higher than found in any of our AGB donor stars. In fact, it is probable that this star did not experience a standard $s$ process and is in fact a member of the CEMP-r/s subclass. Measurement of europium in this object would confirm this. For TY Gru, we obtain a tolerable fit for accretion of around 0.1\ms\ of material from a 1\ms\ companion.

Further analysis of a set of moderate-resolution spectroscopic observations of RR Lyrae CEMP candidates is currently underway.  Such studies, in communion with high-resolution analysis of confirmed CEMP RR Lyrae stars, will enable us to to further constrain the likely progenitors of the CEMP-s class in terms of both the mass of the donor AGB star and the subsequent mixing history of the accreted material.

\section{Acknowledgements}
The authors thank Amanda Karakas for providing the AGB yields in machine-readable format, and Camilla Hansen  and Charles Kuehn for useful discussion. RJS is the recipient of a Sofja Kovalevskaja Award from the Alexander von Humboldt Foundation. He is grateful to RSAA for his Stromlo Fellowship. CRK acknowledges support from Australian Research Council (Super Science Fellowship; FS110200016). TCB acknowledges partial support from grant PHY 08-22648; Physics Frontier Center/Joint Institute for Nuclear Astrophysics (JINA), awarded by the US National Science Foundation.

\bibliography{/Users/richardstancliffe/Work/NewBib}

\label{lastpage}
\end{document}